\documentclass[]{spie}  

\usepackage{amsmath}
\usepackage{graphicx}
\usepackage{mathrsfs}
\usepackage{ntheorem}
\usepackage{mathtools}
\usepackage{enumerate}
\usepackage{enumitem}
\usepackage{physics}
\usepackage{subfigure}
\usepackage{overpic}
\usepackage{epstopdf}
\usepackage{array}
\usepackage{multirow}
\usepackage{enumitem}
\usepackage{array}
\usepackage{booktabs}
\usepackage{fancyhdr}
\usepackage{url}
\usepackage{hyperref}
\usepackage{xcolor}
\hypersetup{colorlinks=true,citecolor=blue,linkcolor=blue,urlcolor=blue,pdfstartview=FitH,bookmarksopen=true}
\usepackage{balance}

\newcommand{\ot}{\otimes}
\newcommand{\bkt}[2]{\langle{#1}|{#2}\rangle}

\title{Real-time analog circuit for auto-correlative weak-value amplification in the time domain}

\author[a]{Jing-Hui Huang}
\author[b]{Guang-Jun Wang}
\author[a,${*}$]{Xiang-Yun Hu}
\affil[a]{Hubei Subsurface Multiscale Image Key Laboratory, School of Geophysics and Geomatics, China University of Geosciences, Lumo Road 388, 430074 Wuhan, China.}
\affil[b]{School of Automation, China University of Geosciences, Wuhan 430074, China}

\authorinfo{Further author information: \\Jing-Hui Huang: E-mail: jinghuihuang@cug.edu.cn;  Xiang-Yun Hu: E-mail: xyhu@cug.edu.cn}

\begin{document} 
\maketitle

\begin{abstract}
The auto-correlative weak-value amplification (AWVA) technique demonstrates distinct advantages over standard weak-value amplification (SWVA) for quantum parameter estimation. To achieve enhanced precision in real-time parameter estimation, the AWVA requires additional resources compared to SWVA, namely real-time multiplication and integrator modules. We implemented a real-time analog circuit for AWVA using an AD835 multiplier and an NE5532 operational amplifier for the integrator. The circuit was tested using Gaussian pointers in the AWVA scheme, exhibiting sufficient sensitivity for Gaussian pointers with frequencies $200\;\mathrm{Hz} \leq f \leq 20\;\mathrm{kHz}$. Compared to SWVA, AWVA achieves higher accuracy and superior robustness against noise at signal-to-noise ratios (SNRs) of $-12\;\mathrm{dB} \leq \mathrm{SNR} \leq -4\;\mathrm{dB}$. Beyond quantum metrology, the circuit is applicable to diverse detection schemes for correlated signals.
\end{abstract}


\section{Introduction}
The implementation of rapid, sensitive, quantitative, low-cost, and real-time parameter estimation and feedback control has garnered significant interest across laboratory-based systems and various fields~\cite{Rossi2018,Hoshi2022}. Such processing can be achieved using either traditional analog circuits or digital circuits such as field programmable gate arrays (FPGAs). 
%
Analog electronic feedback loops have demonstrated particular utility in quantum control applications, notably for stabilizing mechanical quantum systems~\cite{Rossi2018}. Similar techniques have been implemented in large-scale instrumentation such as the AURIGA gravitational wave detector~\cite{PhysRevLett.101.033601}. Meanwhile, custom FPGA systems are increasingly prevalent in sensing and quantum information applications~\cite{Rossi2018,Philips2022}, offering superior computational capabilities and noise immunity. Nevertheless, analog circuits retain relevance due to their low latency, cost-effectiveness, and efficiency for fundamental operations.

Weak measurements have established a robust framework for quantum metrology and information processing~\cite{AAV,RevModPhys.86.307,PhysRevLett.118.070802,PhysRevLett.108.070402,PhysRevX.4.011031,PhysRevLett.125.080501,PhysRevLett.126.220801,Liu:22}. Recent advances include implementations within double-slit interferometers that enhance both sensitivity and signal-to-noise ratio~\cite{PhysRevLett.134.080802}, alongside ongoing refinements of weak measurement techniques~\cite{PhysRevA.102.023701,PhysRevA.100.012109,PhysRevA.105.013718,AWVA}. 
Our group has introduced auto-correlative weak-value amplification (AWVA) to enhance estimation precision under conditions of strong Gaussian white noise~\cite{AWVA}. This technique extends standard weak-value amplification (SWVA)~\cite{AAV} by implementing two simultaneous auto-correlative weak measurements to obtain the auto-correlation intensity $\Theta$. While our initial Simulink simulations~\cite{AWVA} demonstrated the approach using discrete signal processing, they revealed a critical dependence of precision on sampling frequency. Higher sampling frequencies improve precision at the cost of increased computational demands. Consequently, practical AWVA implementation requires either analog circuits or optimized digital systems capable of real-time auto-correlation computation.

Here, we develop a real-time analog circuit implementing the product and integrator operations essential to AWVA. Our design employs an AD835 four-quadrant multiplier, NE5532 dual low-noise operational amplifiers, and an oscilloscope for readout. We characterize the circuit by: (1) validating individual component performance; (2) comparing measured auto-correlation intensities $\Theta(t)$ against theoretical predictions for various input signals; and (3) benchmarking AWVA against SWVA under controlled noise amplitudes. Our results demonstrate the circuit's viability for AWVA applications, exhibiting significant noise robustness.

\section{Auto-correlative weak-value amplification}
\label{Sec:AWVA-princple}
\begin{figure}[htp!]
\centering\includegraphics[width=13.1cm]{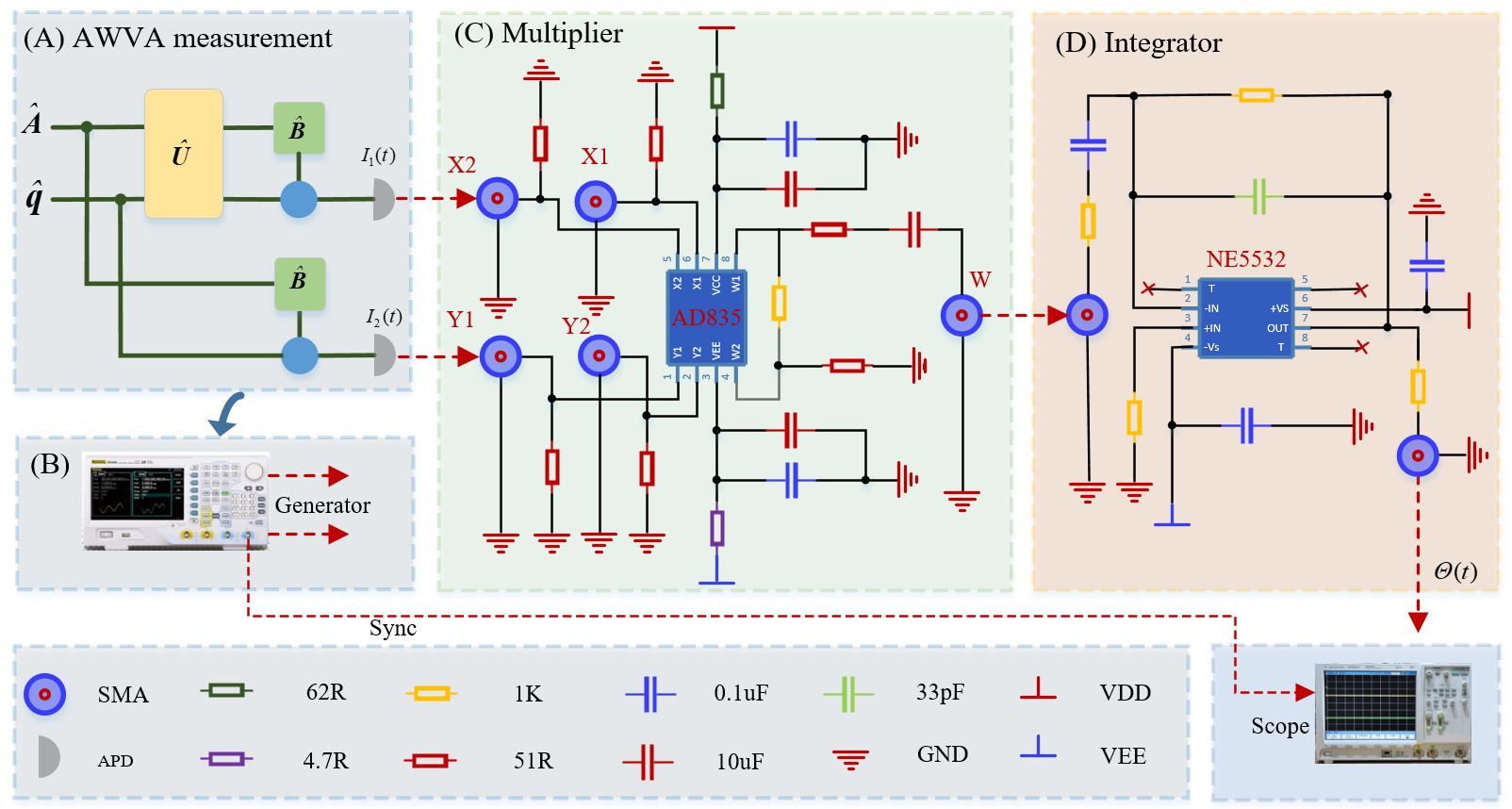}
\caption{\label{Fig:schemeAll} 
Schematic diagram of the system. (A) The AWVA measurement.
(B) An alternative plan for generating autocorrelated signals.
(C) Multiplier diagram with the AD835 chip. 
(D) Integrator diagram with the NE5532 chip. 
}
\end{figure}

As shown in Fig.~\ref{Fig:schemeAll}(A), we present the AWVA scheme in the time domain. AWVA includes one SWVA and another projective measurement without weak coupling. 
In SWM, we prepare the initial state of a two-level system as the polarization state $\ket{\Phi_{i}} = (\ket{H} + \ket{V})/\sqrt{2}$, while the pointer state is prepared in the position basis $\ket{\Psi_{i}} = \int dq f(q) \ket{q}$. Here $\hat{q}$ denotes the longitudinal position along the light's propagation direction in a co-moving frame, with $\ket{H}$ and $\ket{V}$ representing horizontal and vertical polarization states, respectively. The observable operator is $\hat{A} = \ket{H}\bra{H} - \ket{V}\bra{V}$, yielding the joint input state:
$\ket{\Phi_{i}} \otimes \ket{\Psi_{i}} = \frac{1}{\sqrt{2}} (\ket{H} + \ket{V}) \int dq f(q) \ket{q}.$

The system and pointer undergo weak coupling described by the von Neumann Hamiltonian~\cite{AWVA}: $\hat{H} = g(t) \hat{A} \otimes \hat{p},$ where $\hat{p}$ is the momentum operator conjugate to $\hat{q}$ (corresponding to optical frequency in the time/position domain), and $g(t)$ is the instantaneous coupling rate. Solving the time-dependent Schrödinger equation yields the unitary evolution operator:
$\hat{U} = \exp\left(-i \int_{t_i}^{t_f} \hat{H} dt\right) = \exp(-i\gamma \hat{A} \otimes \hat{p}),$ where $\gamma = \int_{t_i}^{t_f} g(t) dt$ defines the coupling strength. 
Following established WVA protocols~\cite{PhysRevA.102.023701,PhysRevA.100.012109,PhysRevA.105.013718}, we use time-delay estimation as our benchmark application, where $\tau \equiv \gamma$ represents the unknown parameter. While such small temporal shifts ($\tau$) typically evade direct projective measurement, post-selected weak measurements enable their precise estimation. To extract $\tau$ from temporal pointer displacements, we post-select the system state as:
$\ket{\Phi_{f}} = \sin\left(-\frac{\pi}{4} + \alpha\right) \ket{H} + \cos\left(-\frac{\pi}{4} + \alpha\right) \ket{V},$
with a projective operator denoted as $\hat{B}=\ket{H} \bra{H}-\ket{V} \bra{V}$.
One then obtains the weak value as:
$A_{w}:=\frac{\bra{\Phi_{f}}\hat{A}\ket{\Phi_{w}}}{\bkt{\Phi_{f}}{\Phi_{w}}} =-{\rm cot} \alpha \, .$
Then, the final state of the pointer can be obtained:
$
\ket{\Psi_{f}}
=\bra{\Phi_{f}}e^{-i\tau\hat{A}\ot \hat{p}} \ket{\Psi_{i}}  \ket{\Phi_{i}} 
\approx
\bkt{\Phi_{f}}{\Phi_{i}}e^{-i\tau A_{w}\hat{p}}\ket{\Psi_{i}}\, .
$
Normally, the profile of the initial pointer is chosen as the Gaussian profile: $\left|\left\langle t | \Psi_{i}\right\rangle\right|^{2}=I_{0}\frac{1} {(2 \pi \omega^{2})^{1/4}}  e^{-(t-t_{0})^{2}/4\omega^{2}}$. Where $I_{0}$ represents the {normalization} factor and $\omega$ represents the pointer spread.
Finally, the signal $I_{1}(t)$ of the SWVA detected by an APD is given by:
\begin{small}
\begin{eqnarray}
\label{Eq:schme1inter_peobe_final}
I_{1}(t;\tau)= |\bkt{\Phi_{f}}{\Phi_{i}}|^{2} e^{-2i\tau A_{w}\hat{p}} \left|\left\langle t | \Psi_{i}\right\rangle\right|^{2} 
\approx  \frac{I_{0}}{2}\frac{({\rm sin}\alpha)^{2}} {(2 \pi \omega^{2})^{1/4}}   e^{-(t-t_{0}-\delta t)^{2}/4\omega^{2}}.
\end{eqnarray} 
\end{small}
Eq.~(\ref{Eq:schme1inter_peobe_final}) indicates that the time shift $\tau$ can be  amplified and obtained from the shift $\delta t=|\tau {\rm Re}\,[ A_{w}]|= \tau  {\rm cot} \alpha$ of the pointer. At the same time, the signal $I_{2}(t;\tau)$ corresponds to the results of the projective measurement without weak coupling ($\hat{U}$) is given by:
\begin{small}
\begin{eqnarray}
\label{Eq:schme1inter_peobe_final22}
I_{2}(t;)= |\bkt{\Phi_{f}}{\Phi_{i}}|^{2}  \left|\left\langle t | \Psi_{i}\right\rangle\right|^{2} 
\approx  \frac{I_{0}}{2}\frac{({\rm sin}\alpha)^{2}} {(2 \pi \omega^{2})^{1/4}}   e^{-(t-t_{0})^{2}/4\omega^{2}}.
\end{eqnarray} 
\end{small}
In our previous work~\cite{AWVA}, our simulation has demonstrated that the measurement of the new quantity, namely the auto-correlation intensity $\rm \Theta(t)$  outperforms the direct measurement of the pointer shift $I_{1}(t;\tau)$ with smaller statistical errors under strong Gaussian white noise background.  $\rm \Theta(t)$ is calculated by:
\begin{small}
\begin{eqnarray}
\label{Eq:ACIdefine}
{\rm \Theta}_{}(t;\tau)=\int_{0}^{t} I_{1}^{}(t^{\prime};\tau) \times  I_{2}^{}(t^{\prime}) dt^{\prime} 
=  \frac{I_{0}^{2}}{4} \frac{({\rm sin}\alpha)^{4}} {(2 \pi \omega^{2})^{1/8} }   \int _{0}^{t} e^{-[(t^{\prime}-t_{0})^{2}+(t^{\prime}-t_{0}-\delta t)^{2}]/4\omega^{2}} dt^{\prime}\, . 
\end{eqnarray}
\end{small}
%

\section{Experimental results and Discussion} \label{Sec:results}
The experimental setup is shown in Fig.~\ref{Fig:schemeAll} and Fig.~1(B) of the \href{https://doi.org/10.7910/DVN/EOYJHO,}{Supplemental Material}\cite{DVN/EOYJHO_2025}. Key components include: A function/arbitrary waveform generator (RIGOL DG4202, 200 MHz, 500 MSa/s) for signal generation; A digital oscilloscope (Agilent DSO7052B, 500 MHz, 4 GSa/s) for signal acquisition and analysis.
The time delay $\tau$ (the estimation parameter in AWVA) manifests as a temporal shift $\delta t$ in the pointer. We implement $\delta t$ by phase-shifting one input signal using the generator. This approach substitutes the full AWVA measurement process (Fig.~\ref{Fig:schemeAll}A) with direct generation of the weak-value-modified signal $I_1(t;\tau)$ and reference signal $I_2(t)$ (Fig.~\ref{Fig:schemeAll}B). The signal $\Theta(t)$ is then computed in real-time by the analog circuit and displayed on the oscilloscope.

Before AWVA implementation, we characterized the multiplier and integrator circuits individually (see \href{https://doi.org/10.7910/DVN/EOYJHO,}{Supplemental Material} Sec.~II and Sec.~III). These measurements yield the overall gain coefficient $\Gamma = \Gamma_M \times \Gamma_I$. For Gaussian pointers, we obtain $\Gamma^{ga} = \Gamma_M^{ga} \times \Gamma_I^{ga} = 1.87 \times 4470 = 8358$, representing significant signal amplification relative to theoretical predictions. This gain enhancement is particularly valuable since post-selection in AWVA reduces the success probability~\cite{PhysRevLett.112.040406}. Noise characterization (\href{https://doi.org/10.7910/DVN/EOYJHO,}{Supplemental Material} Sec. IV) confirms minimal Gaussian noise impact on AWVA performance for pointer frequencies of 200 Hz and 2000 Hz.

\begin{figure}[htp!]
	\centering
{
	\begin{minipage}{0.48\linewidth}
\centerline{\includegraphics[scale=0.14,angle=0]{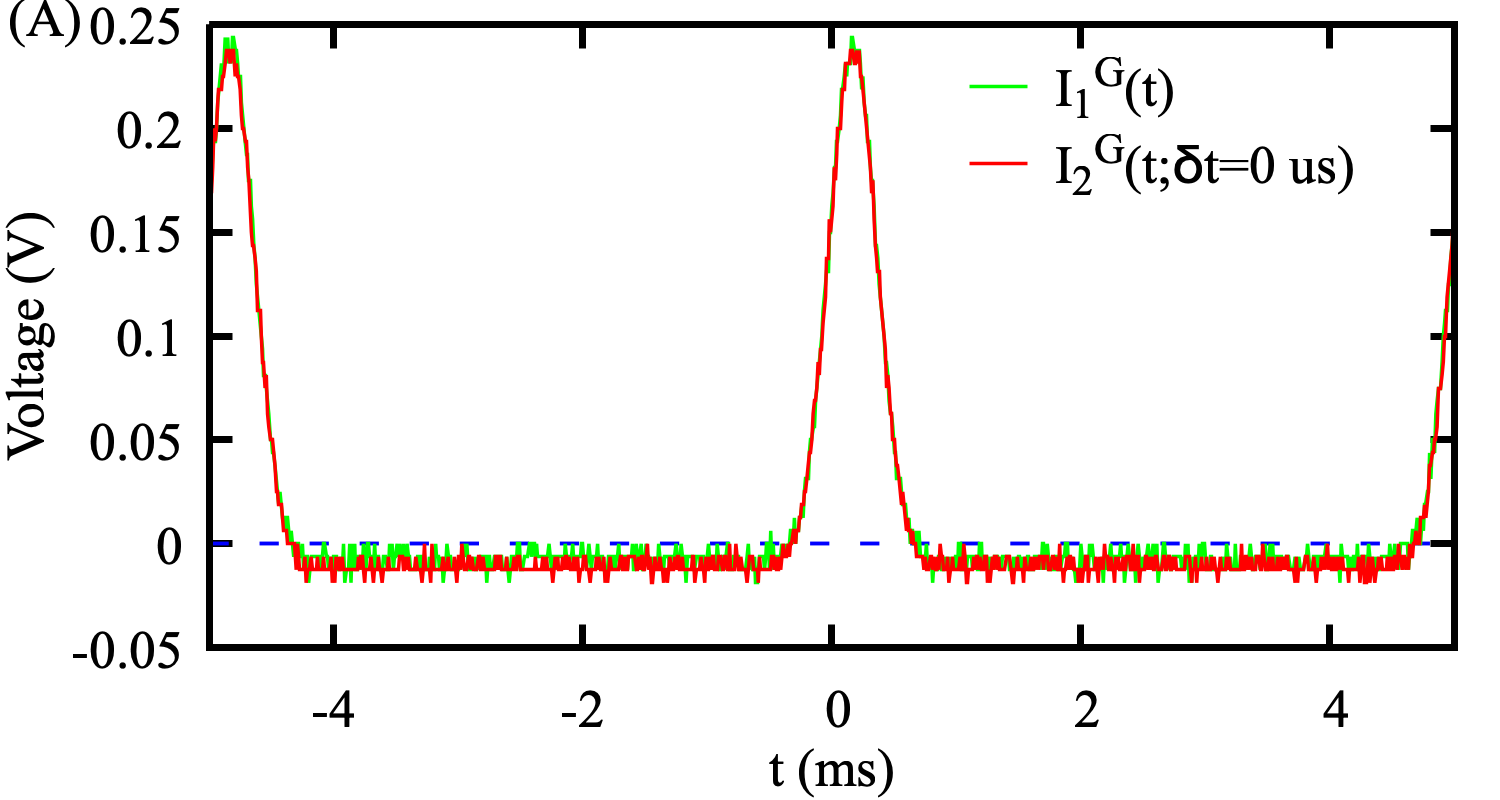}}
	\end{minipage}
}
{
	\begin{minipage}{0.48\linewidth}
	\centering	\centerline{\includegraphics[scale=0.14,angle=0]{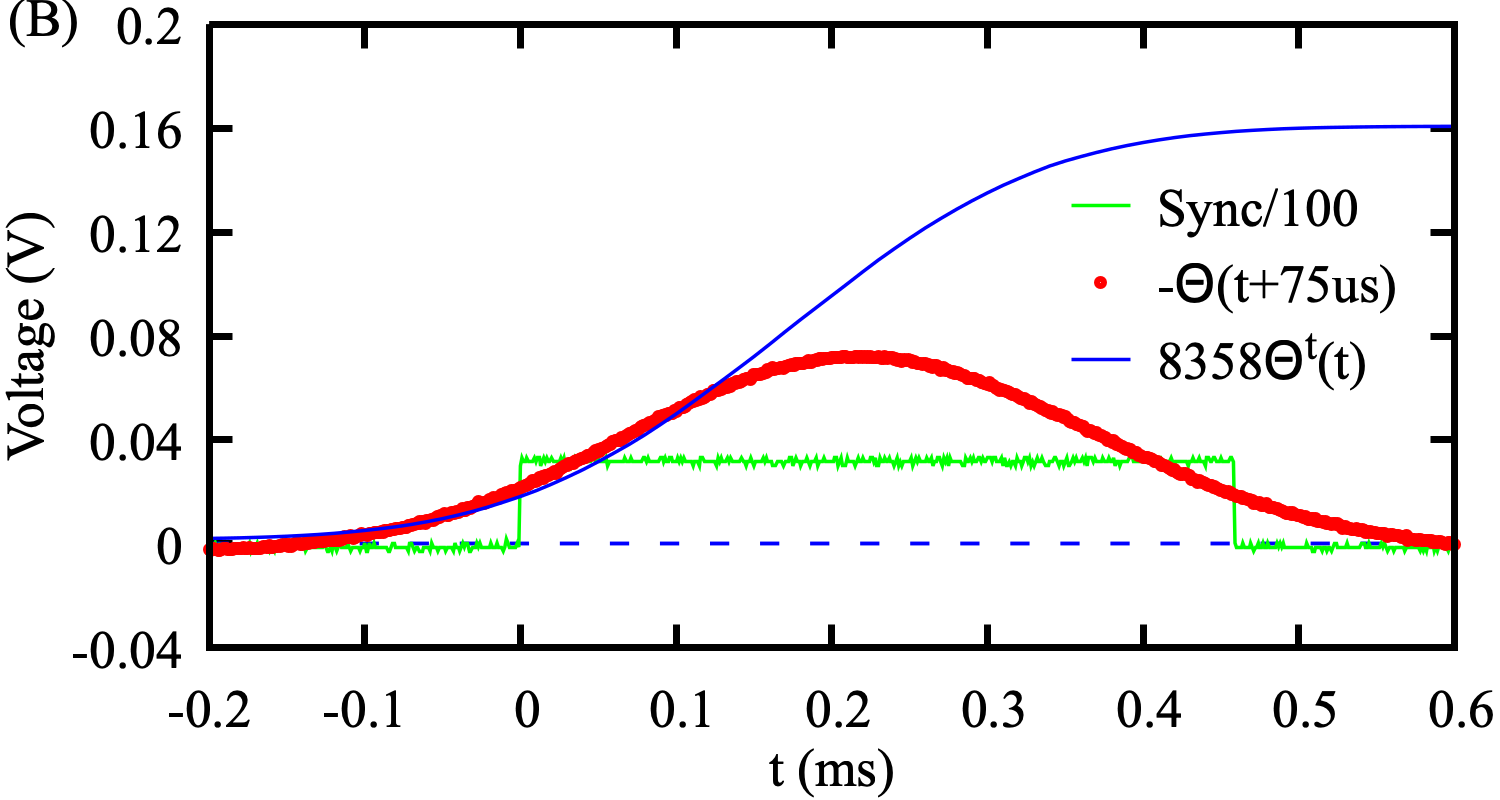}}
	\end{minipage}
}
\caption{\label{Fig:compareExperimrntAndTheory}
(A) The input of signals $I^G_1(t)$ and $I^G_2(t;\delta t=0 us)$. (B) The theoretical $ \Theta^t(t)$ and the experimental $ \Theta(t)$.
}
\end{figure}

We next quantify $\Theta(t)$ for the Gaussian pointer in noise-free conditions. The input signals are defined as:
\begin{small}
\begin{align}
\label{Eq:multiplier_Gauss_result1}
I^G_1(t) = 0.248 \exp\left[-2 \left(\frac{t - 1.71 \times 10^{-4}}{3.88 \times 10^{-4}}\right)^2\right] - 0.01,
I^G_2(t) = 0.248 \exp\left[-2 \left(\frac{t - 1.71 \times 10^{-4} - \delta t}{3.88 \times 10^{-4}}\right)^2\right] - 0.01
\end{align}
\end{small}
with $\delta t$ representing the temporal shift between signals. Figure~\ref{Fig:compareExperimrntAndTheory} compares these inputs with the theoretical auto-correlation $\Theta^t(t; t_0)$ at $\delta t = 0$. 
Figure~\ref{Fig:compareExperimrntAndTheory}(B) shows excellent agreement between scaled theory ($8358\Theta^t(t)$) and phase-corrected experiment ($-\Theta(t + 75 \mu s)$). The $75 \mu s$ phase lag originates from circuit capacitance and was determined by curve alignment. The curves coincide within $-0.2 \leq t \leq 0.1$ ms (half-cycle duration), validating our circuit implementation. In this paper, the phase lag $t_\phi$ does not affect sensitivity $K$, defined as:
\begin{small}
\begin{equation}
\label{Eq:sensitivity-defined}
K = -\frac{\Theta(t + t_\phi; \delta t) - \Theta(t + t_\phi)}{\delta t} = -\frac{\Theta(t; \delta t) - \Theta(t)}{\delta t}
\end{equation}
\end{small}

Figure~3 of \href{https://doi.org/10.7910/DVN/EOYJHO,}{Supplemental Material} shows single $\Theta(t;\delta t)$ measurements for different time shifts $\delta t$ in signals $I^G_1(t)$ and $I^G_2(t)$ at various frequencies $f$. The $\Theta(t;\delta t)$ profile exhibits significant frequency dependence. Table~\ref{Table:sensitivityDifferentFreq} summarizes sensitivity measurements across frequencies, where maximum values $\text{Max}[\Theta(t;\delta t)]$ and $\text{Max}[\Theta(t)]$ were directly recorded from oscilloscope amplitudes with associated standard deviations. 
Key observations from Table~\ref{Table:sensitivityDifferentFreq} include: (i) Sensitivity $K$ increases while its standard error decreases with larger $\delta t$, demonstrating enhanced precision for greater time shifts.
(ii) Higher frequencies $f$ increase sensitivity (and thus precision) at the cost of reduced $\Theta(t)$ amplitude, limited by the NE5532's performance.
(iii) Oscilloscope measurements yield stable mean values for $\text{Max}[\Theta(t;\delta t)]$ and $\text{Max}[\Theta(t)]$ at $C > 10,000$, validating mean $K$ as a reliable metric despite its standard error.

\begin{table}[htb!]
\centering
\setlength{\tabcolsep}{8.0mm}
\renewcommand{\arraystretch}{0.9}
{
\caption{
\label{Table:sensitivityDifferentFreq}
Characteristic results for measuring $\delta t$ (us) with different frequency f (Hz). The maximum values  Max[$\Theta$(t;$\delta$t)] (mV) and $\rm Max[\Theta(t)]$ (mV)  are directly read out as the amplitude with their standard deviation from the scope. The unite of $K$ is $10^{-2}$ mV/us. 
}
\begin{tabular}{rrrrr}
\hline
\hline
 f                 
& $\delta t$
& $\rm Max[\Theta(t)]$ 
& Max[$\Theta$(t;$\delta$t)] 
& $K$
\\ \hline
200 
 & 50
 & $76.797  \pm 0.839 $             
 & $75.978  \pm 0.847 $               
 & $1.638 \pm 3.372 $
\\
200
 & 100 
 & $76.797  \pm 0.839 $             
 & $72.278  \pm 0.879 $               
 & $4.519 \pm 1.718 $
\\ \hline
2000 
 & 5 
 & $28.732  \pm 0.219 $                      
 & $28.273  \pm 0.241 $               
 & $9.180 \pm 9.200 $
\\
2000 
 & 10 
 & $28.732  \pm 0.219 $                      
 & $26.837  \pm 0.178 $               
 & $18.950   \pm 3.970 $
\\ \hline
20000
 & 0.5 
 & $6.8856  \pm 0.120 $                
 & $6.8087  \pm 0.135 $               
 & $15.380   \pm 50.10 $
\\
20000 
 & 1.0 
 & $6.8856  \pm 0.120 $
 & $6.4790  \pm 0.113 $             
 & $40.660   \pm 23.33 $
\\ 
\hline
\hline
\end{tabular}
}
\end{table}

\begin{figure}[htp!]
	\centering
{
	\begin{minipage}{0.48\linewidth}
	\centering
\centerline{\includegraphics[scale=0.14,angle=0]{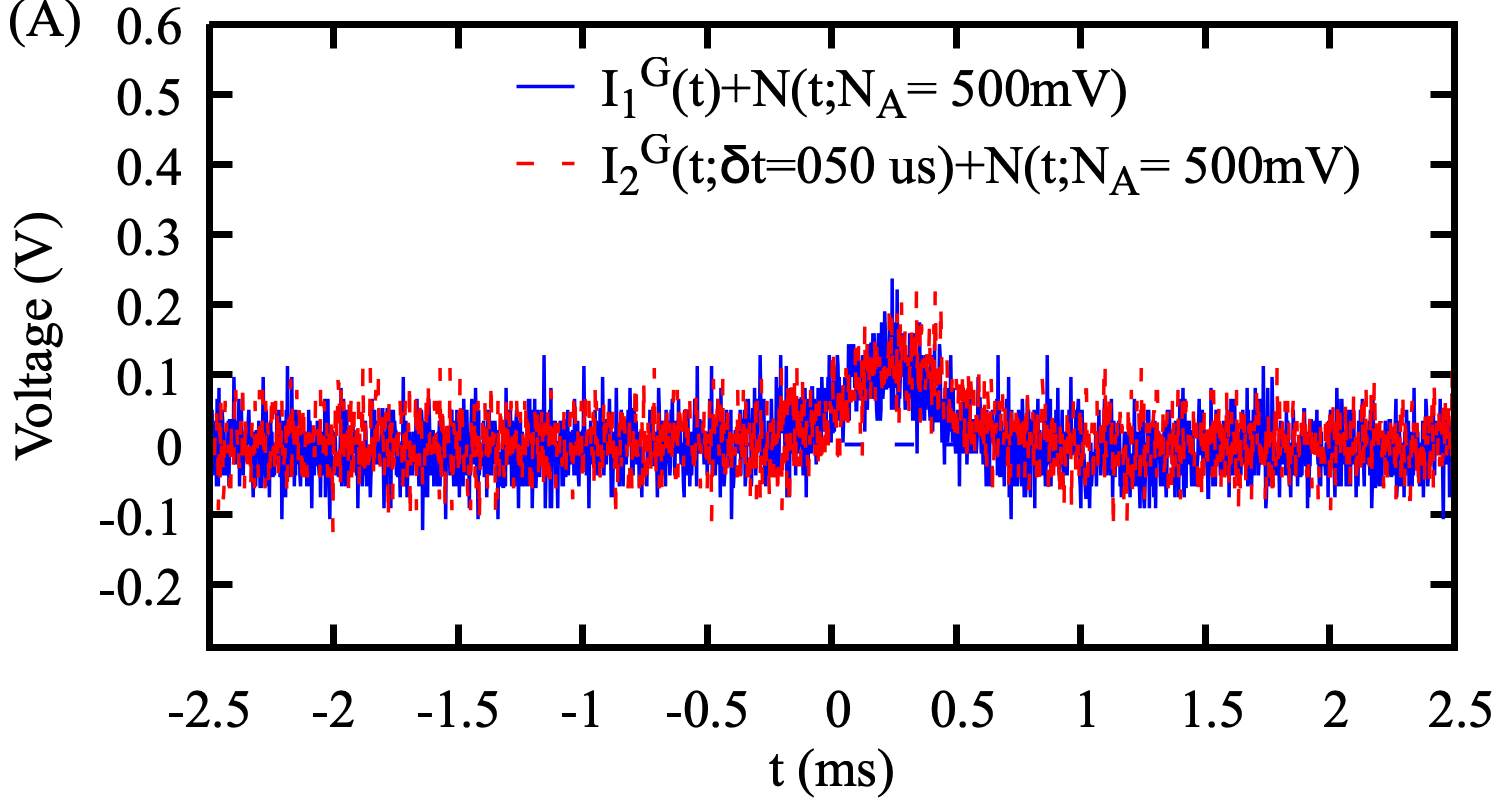}}
	\end{minipage}
}
{
	\begin{minipage}{0.48\linewidth}
	\centering
	\centerline{\includegraphics[scale=0.14,angle=0]{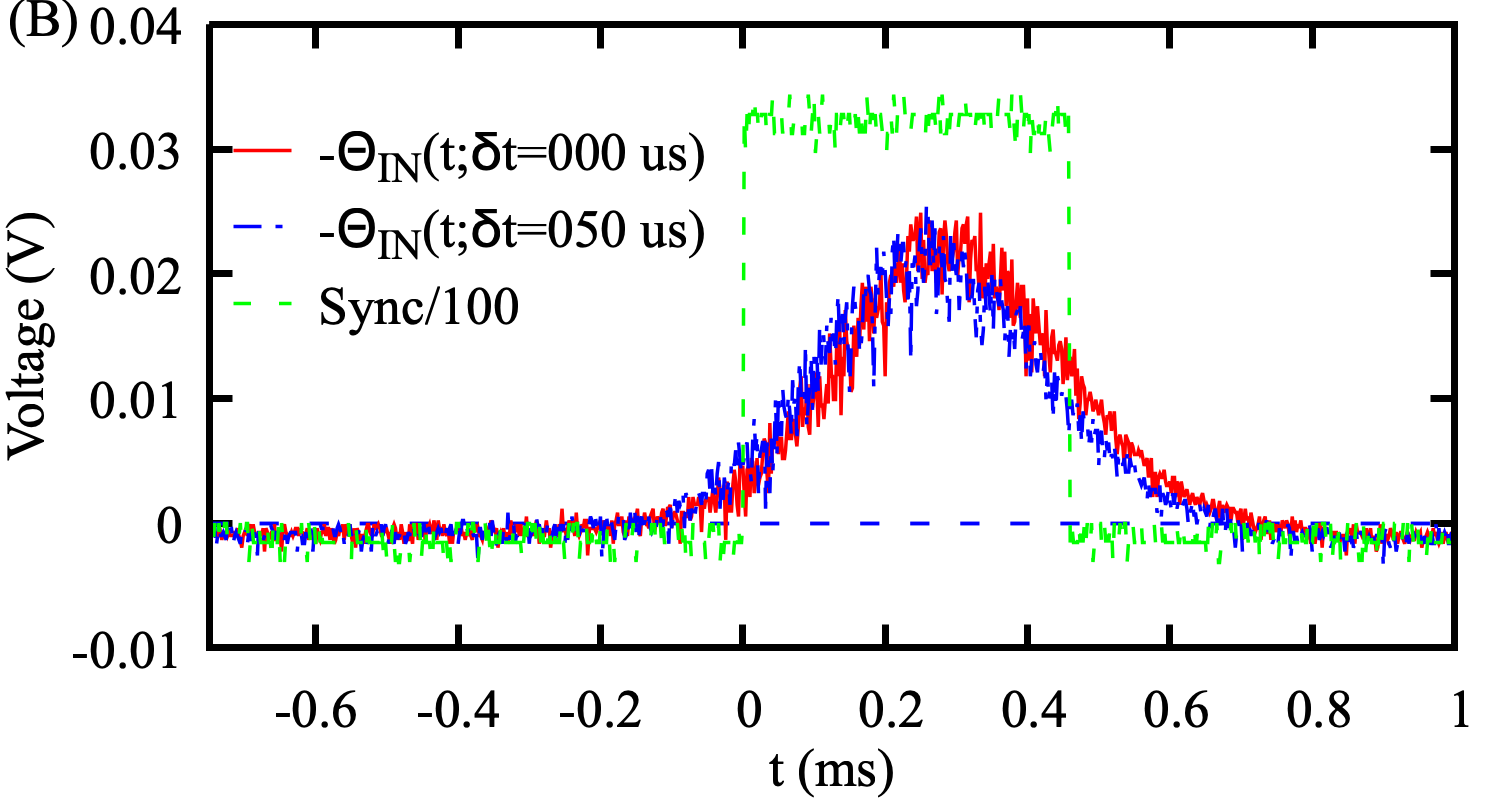}}
	\end{minipage}
}

{
	\begin{minipage}{0.48\linewidth}
	\centering
	\centerline{\includegraphics[scale=0.14,angle=0]{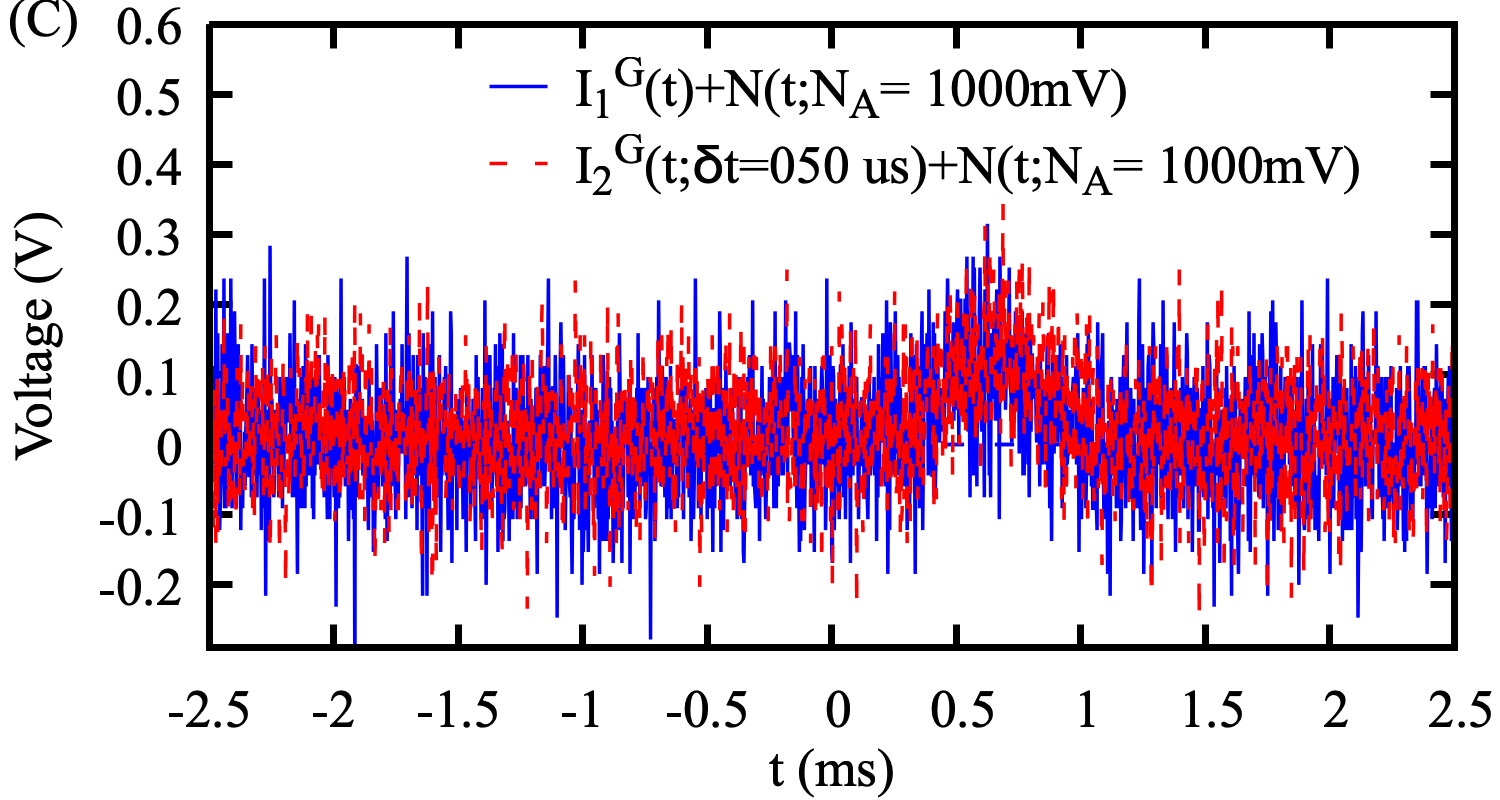}}
	\end{minipage}
}
{
	\begin{minipage}{0.48\linewidth}
	\centering
	\centerline{\includegraphics[scale=0.14,angle=0]{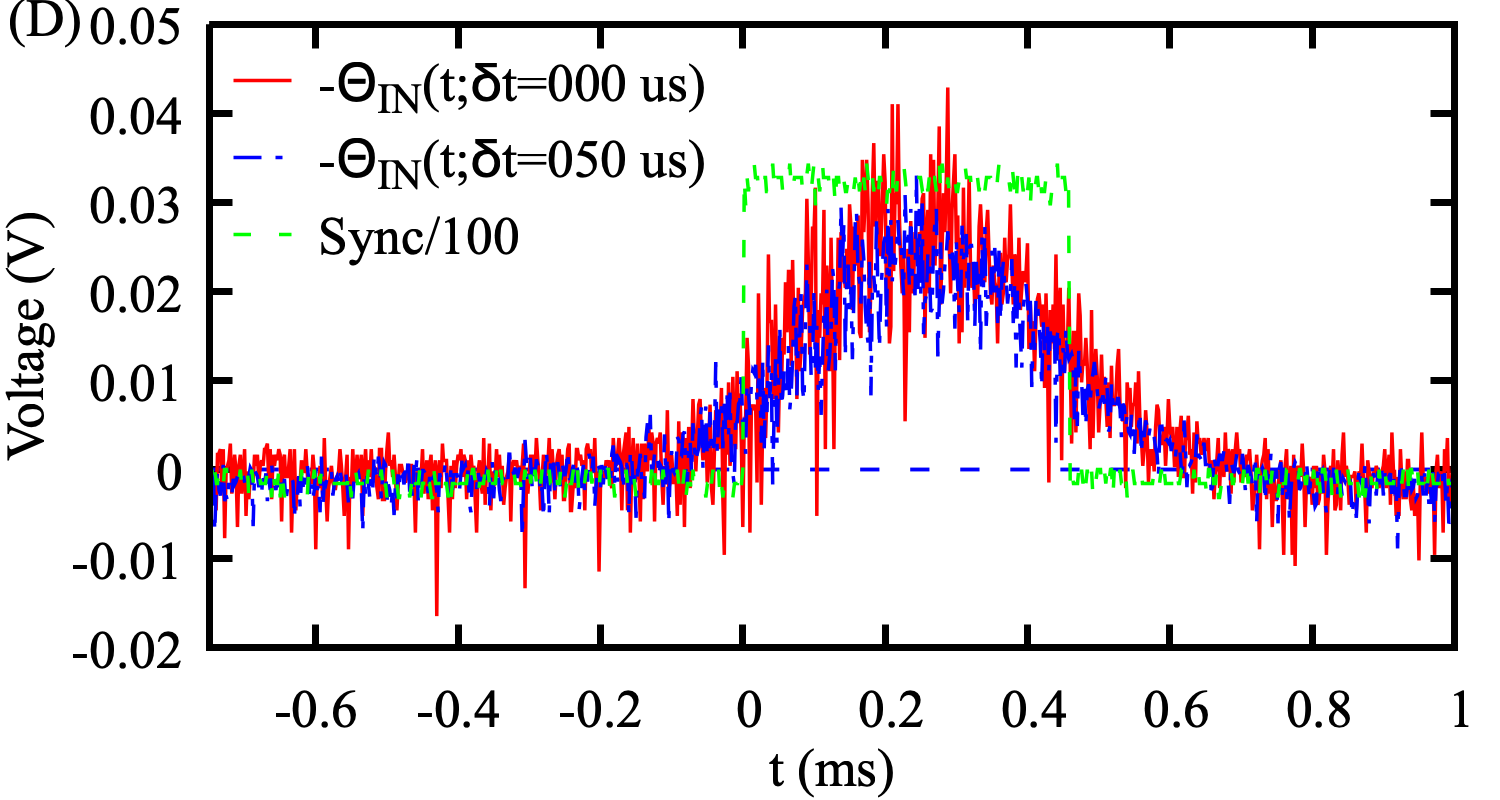}}
	\end{minipage}
}
\caption{\label{Fig:result-different-noiseSNR}
AWVA measurements under strong Gaussian noise. (A, C) Input signals $I^N_1(t)$ (solid) and $I^N_2(t; \delta t = 50 \mu s)$ (dashed) at noise amplitudes (A) $N_A = 500$ mV and (C) $N_A = 1000$ mV. (B, D) Corresponding auto-correlation outputs $-\Theta(t)$ showing (B) 500 mV and (D) 1000 mV noise responses.
}
\end{figure}

Finally, we characterize AWVA performance under strong Gaussian noise by injecting noise $\mathbf{N}(t, N_A)$ into the pointer signals. The resulting auto-correlation intensity $\Theta_{\mathrm{IN}}(t; \delta t)$ is defined as:
\begin{equation}
\label{Eq:ACIdefineWihtnoiseAndSignal}
\Theta_{\mathrm{IN}}(t; \delta t) = \int_{0}^{t} \left[I_{1}^{G}(t'; \delta t) + \mathbf{N}(t', N_A)\right] \times \left[I_{2}^{G}(t') + \mathbf{N}(t', N_A)\right] dt',
\end{equation}
where the pointer amplitude is fixed at 250 mV and the noise amplitude $N_A$ ranges from 0 to 2000 mV. Figure~\ref{Fig:result-different-noiseSNR} shows representative single-shot measurements, while statistical results (mean and standard deviation over $>10^4$ trials) are tabulated in Table~1 of the \href{https://doi.org/10.7910/DVN/EOYJHO,}{Supplemental Material}.
To quantify noise robustness, we compute normalized sensitivities for both techniques:
\begin{align}
\label{Eq:defineSensitivity1}
K^{W} = \frac{\delta t^{\mathrm{Scope}}}{\delta t} = \frac{\delta t^{\mathrm{Scope}}}{50  \mu\text{s}}, 
K^{A} = \frac{K(N_{A})}{K(N_{A} =0  \text{mV})} = \frac{K(N_{A})}{1.208 \times 10^{-2}  \text{mV}/\mu\text{s}},
\end{align}
where $\delta t^{\mathrm{Scope}}$ is the time delay measured by the scope. Following Ref.~\cite{AWVA}, we define the signal-to-noise ratio as:
\begin{equation}
\mathrm{SNR} = 20 \log_{10} \left( \frac{\max |I_{1}^{\mathrm{out}}(t)|}{\max |\mathbf{N}(t)|} \right).
\end{equation}

\begin{small}
\begin{table}[t]
\centering
\setlength{\tabcolsep}{
6.5mm}
\renewcommand{\arraystretch}{0.9}
{
\caption{\label{Table:sensitivityDifferentSNR_normalized}
{ The normalized sensitivities $K^{S}$ in SWVA and $K^{A}$ in AWVA for measuring $\delta t$= 50 us at frequency f= 200 Hz under Gaussian noises with different amplitude  $N_A$ (mV). 
}}
\begin{tabular}{rrr|rrrr}
\hline
\hline
 $N_A$
& $K^{S}$
& $K^{A}$
& $N_A$
& $K^{S}$
& $K^{A}$
\\ \hline
 $2$
 & $1.00  \pm 0.09 $ 
 & $1.00  \pm 0.48$             
 & 200            
 & $1.04  \pm 0.46 $ 
 & $1.04  \pm 0.94$  
\\ 
 10
 & $1.02  \pm 0.04 $ 
 & $0.95  \pm 0.51 $             
 & 300            
 & $0.88  \pm 0.94 $ 
 & $0.98  \pm 1.25$  
\\ 
 20
 & $1.02 \pm 0.07 $ 
 & $0.94  \pm  0.59 $             
 & 400            
 & $0.66  \pm 2.90 $ 
 & $1.05  \pm 1.44$  
\\ 
 30
 & $1.04   \pm 0.08 $ 
 & $1.01 \pm  0.71 $             
 & 500            
 & $2.02  \pm 10.9 $ 
 & $1.11  \pm 1.80$  
\\ 
 40
 & $ 1.00  \pm 0.09 $ 
 & $0.93  \pm 0.62 $             
 & 600            
 & $2.24  \pm 12.1 $ 
 & $1.32  \pm 1.89$  
\\ 
 50
 & $1.04  \pm 0.13 $ 
 & $0.94  \pm  0.58$             
 & 700            
 & $4.10  \pm 13.4 $ 
 & $1.22  \pm 2.24$  
\\ 
 60
 & $ 0.98  \pm 0.14 $ 
 & $0.90  \pm  0.53 $             
 & 800            
 & $0.60 \pm 9.40 $ 
 & $1.30  \pm 2.55$  
\\ 
 70
 & $ 1.06  \pm 0.22 $ 
 & $0.80 \pm  0.62 $             
 & 900            
 & $0.32  \pm 8.00 $ 
 & $0.91  \pm 2.91$  
\\ 
 80
 & $ 1.20  \pm  0.28$ 
 & $1.00  \pm  0.61 $             
 & 1000            
 & $0.46  \pm 6.28 $ 
 & $1.72  \pm 3.52$  
\\ 
 90
 & $1.24  \pm 0.30 $ 
 & $0.85  \pm  0.65$             
 & 1200            
 & $0.12  \pm 5.76 $ 
 & $1.46  \pm 5.70$  
\\ 
100
 & $0.90  \pm 0.44 $ 
 & $1.04  \pm 0.70$            
 & 1400            
 & $-0.50  \pm 6.90 $ 
 & $1.49  \pm 8.18$  
\\ 
 & 
 &           
 & 1600            
 & $0.36  \pm 5.46 $ 
 & $1.21  \pm 11.9$  
\\ 
 & 
 &           
 & 1800            
 & $-1.86  \pm 7.68$ 
 & $1.21  \pm 18.3$  
\\ 
 & 
 &            
 & 2000            
 & $-1.22  \pm 7.38 $ 
 & $1.12  \pm 26.6$  
 \\
\hline
\hline
\end{tabular}
}
\end{table}
\end{small}

\begin{figure}[htb!]
	\centering
{
	\begin{minipage}{0.48\linewidth}
	\centering
	\centerline{\includegraphics[scale=0.14,angle=0]{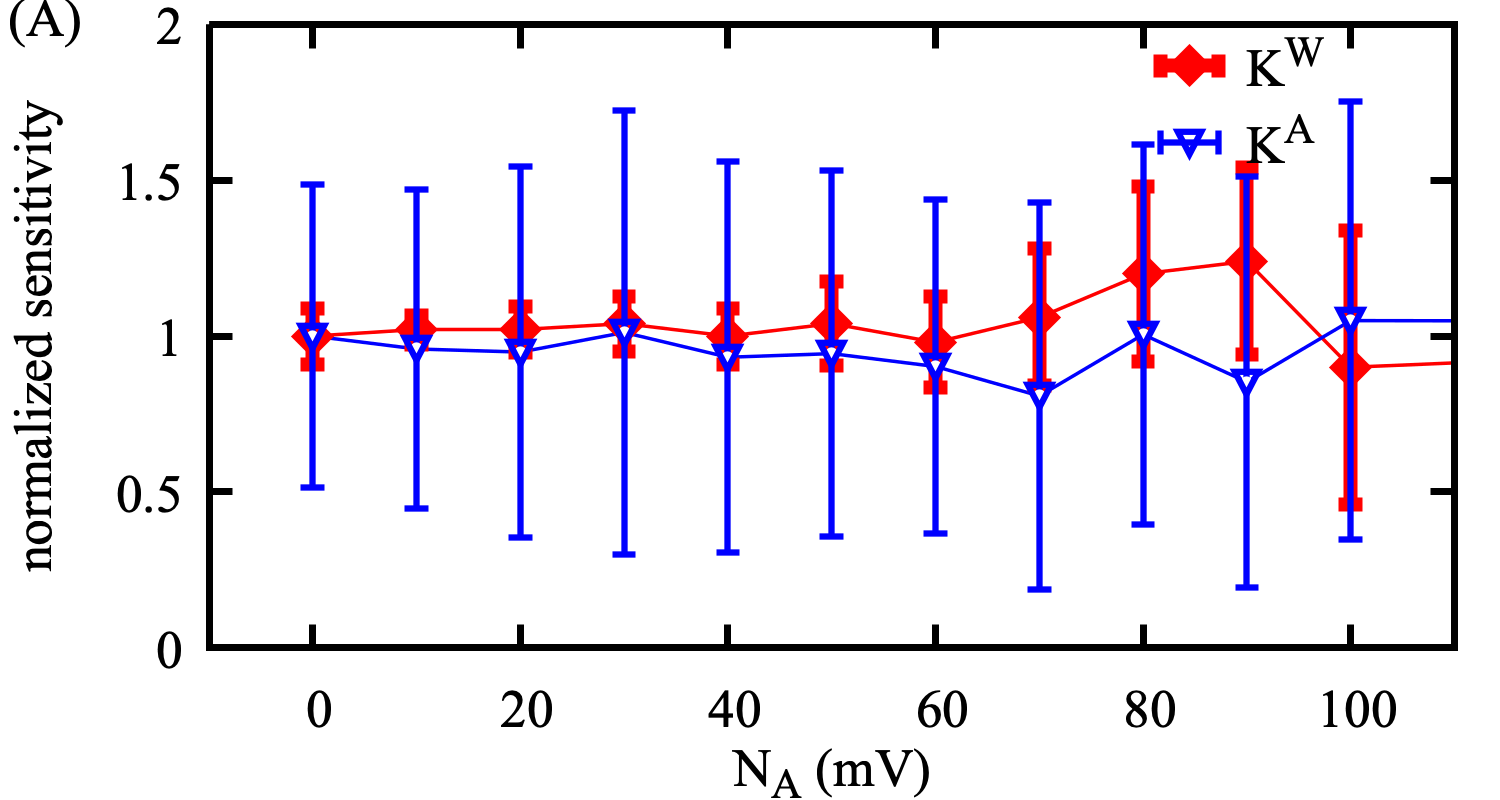}}
	\end{minipage}
}
{
	\begin{minipage}{0.48\linewidth}
	\centering
	\centerline{\includegraphics[scale=0.14,angle=0]{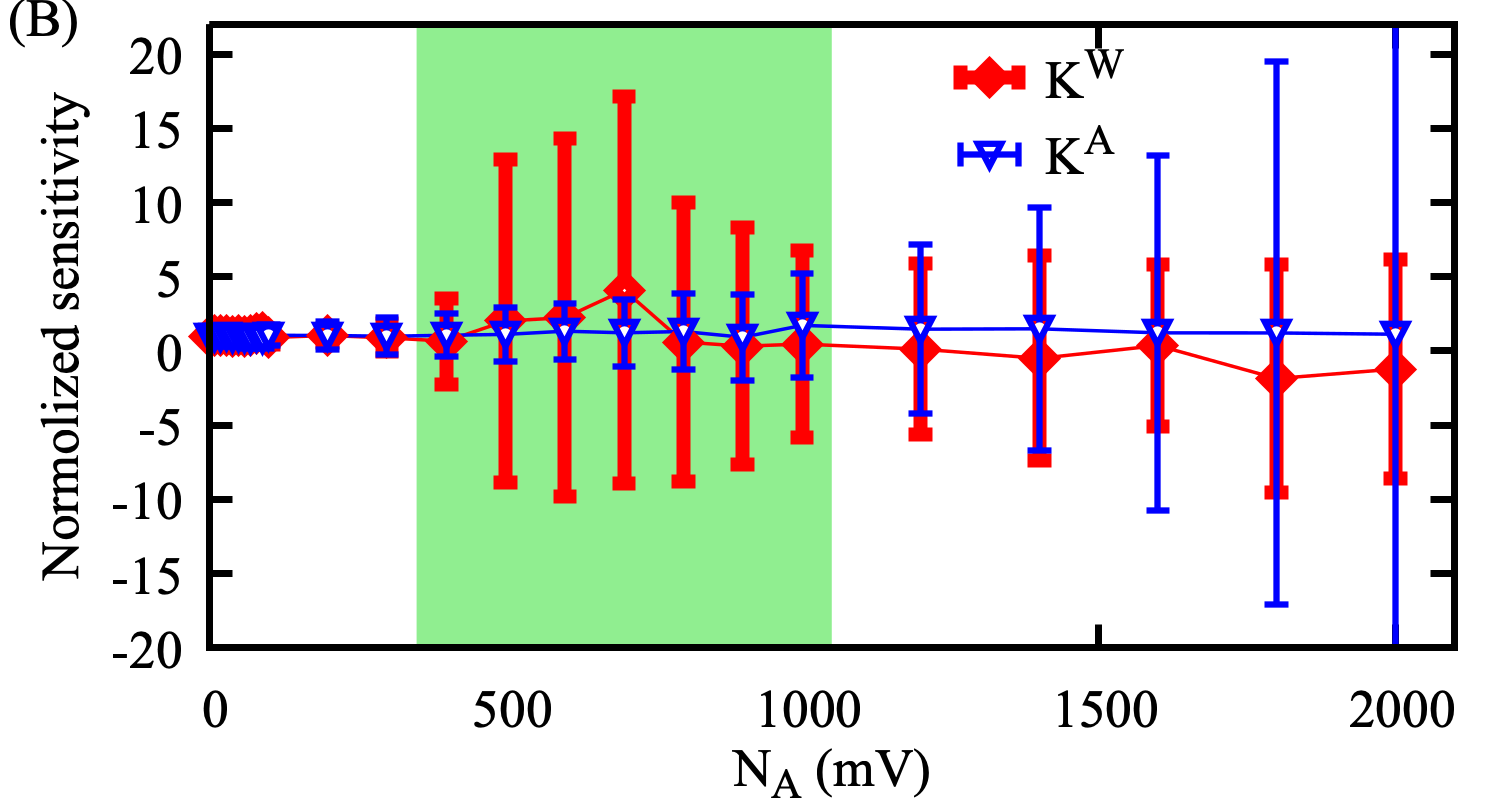}}
	\end{minipage}
}
\caption{\label{Fig:result-different-sensitivity12}
Sensitivities $K^W$ (SWVA) and $K^A$ (AWVA) versus noise amplitude for $\delta t = 50 \mu$s time shifts. The green band highlights the 400-1000 mV range where AWVA achieves smaller error bars than SWVA.
}
\end{figure}

Comparing SWVA and AWVA performance reveals key insights: 
(i) \textbf{Low-noise regime (0 mV $\leq N_A \leq$ 100 mV, SNR $>$ 8 dB)}: SWVA exhibits smaller error bars than AWVA, demonstrating no AWVA advantage under low-noise conditions.
(ii) \textbf{Moderate-noise regime (400 mV $\leq N_A \leq$ 1000 mV, -12 dB $\leq$ SNR $\leq$ -4 dB)}: AWVA achieves superior performance with reduced mean-value deviation and smaller error bars in $K^A$ compared to $K^W$ (Table~\ref{Table:sensitivityDifferentSNR_normalized}, Fig.~\ref{Fig:result-different-sensitivity12}B).
(iii) \textbf{High-noise regime (1200 mV $\leq N_A \leq$ 2000 mV, -18 dB $\leq$ SNR $\leq$ -13 dB)}: While AWVA shows larger $K^A$ error bars than SWVA's $K^W$, it maintains smaller mean-value deviation. Crucially, SWVA yields invalid negative sensitivities at 1400/1800/2000 mV, while AWVA preserves valid positive values.
(iv) \textbf{Noise-scaling effects}: Increasing $N_A$ amplifies both $\Theta_{\mathrm{IN}}$ amplitude and $K^A$ statistical error. This error stems from circuit noise response (Supplemental Material Sec. IV) and represents a systematic component addressable through increased bandwidth.
(v) \textbf{Robustness to unknown noise}: For background noise $2 \leq N_A \leq 2000$ mV, $K^W$ varies over [-1.86, 2.24] (encompassing invalid values), while $K^A$ remains within [0.80, 1.72]. AWVA thus demonstrates superior robustness against unknown noise amplitudes.
\section{Conclusions} \label{Sec:Summary}

In summary, we implemented a real-time analog circuit for auto-correlative weak-value amplification (AWVA) using an AD835 multiplier and NE5532-based integrator. Our hardware-focused approach enables direct computation of the auto-correlation integral. Key findings include:
(i) The NE5532 amplifier provides significant signal gain (up to 8358× for Gaussian pointers), compensating for AWVA's post-selection losses.
(ii)  Experimental validation confirms AWVA's advantage over standard weak-value amplification (SWVA) under moderate-to-high noise (200 mV $\leq N_A \leq$ 2000 mV).
(iii)  The circuit architecture extends beyond quantum metrology to general auto-correlation signal processing.
These results establish a hardware pathway for noise-resilient parameter estimation in real-time sensing applications.

\begin{acknowledgments}
This study was supported by the National Natural Science Foundation of China (Grants No.~42327803 and No.~42488201). 
J-H. Huang acknowledges support from the Hubei Provincial Natural Science Foundation of China (Grant No.~20250650025), the Postdoctoral Fellowship Program and China Postdoctoral Science Foundation under Grant Number BX20250161. 
\end{acknowledgments} 

\bibliography{report} 

\begin{thebibliography}{10}

\bibitem{Rossi2018}
Rossi, M., Mason, D., Chen, J., Tsaturyan, Y., and Schliesser, A., ``{Measurement-based quantum control of mechanical motion},'' {\em Nature}~{\bf 563}(7729),  53--58 (2018).

\bibitem{Hoshi2022}
Hoshi, I., Shimobaba, T., Kakue, T., and Ito, T., ``{Real-time single-pixel imaging using a system on a chip field-programmable gate array},'' {\em Scientific Reports}~{\bf 12}(1),  14097 (2022).

\bibitem{PhysRevLett.101.033601}
Vinante, A., Bignotto, M., Bonaldi, M., Cerdonio, M., Conti, L., Falferi, P., Liguori, N., Longo, S., Mezzena, R., Ortolan, A., Prodi, G.~A., Salemi, F., Taffarello, L., Vedovato, G., Vitale, S., and Zendri, J.-P., ``Feedback cooling of the normal modes of a massive electromechanical system to submillikelvin temperature,'' {\em Phys. Rev. Lett.}~{\bf 101},  033601 (Jul 2008).

\bibitem{Philips2022}
Philips, S. G.~J., M{\c{a}}dzik, M.~T., Amitonov, S.~V., de~Snoo, S.~L., Russ, M., Kalhor, N., Volk, C., Lawrie, W. I.~L., Brousse, D., Tryputen, L., Wuetz, B.~P., Sammak, A., Veldhorst, M., Scappucci, G., and Vandersypen, L. M.~K., ``{Universal control of a six-qubit quantum processor in silicon},'' {\em Nature}~{\bf 609}(7929),  919--924 (2022).

\bibitem{AAV}
Aharonov, Y., Albert, D.~Z., and Vaidman, L., ``How the result of a measurement of a component of the spin of a spin-1/2 particle can turn out to be 100,'' {\em Phys. Rev. Lett.}~{\bf 60},  1351--1354 (Apr 1988).

\bibitem{RevModPhys.86.307}
Dressel, J., Malik, M., Miatto, F.~M., Jordan, A.~N., and Boyd, R.~W., ``Colloquium: Understanding quantum weak values: Basics and applications,'' {\em Rev. Mod. Phys.}~{\bf 86},  307--316 (Mar 2014).

\bibitem{PhysRevLett.118.070802}
Harris, J., Boyd, R.~W., and Lundeen, J.~S., ``Weak value amplification can outperform conventional measurement in the presence of detector saturation,'' {\em Phys. Rev. Lett.}~{\bf 118},  070802 (Feb 2017).

\bibitem{PhysRevLett.108.070402}
Lundeen, J.~S. and Bamber, C., ``Procedure for direct measurement of general quantum states using weak measurement,'' {\em Phys. Rev. Lett.}~{\bf 108},  070402 (Feb 2012).

\bibitem{PhysRevX.4.011031}
Jordan, A.~N., Mart\'{\i}nez-Rinc\'on, J., and Howell, J.~C., ``Technical advantages for weak-value amplification: When less is more,'' {\em Phys. Rev. X}~{\bf 4},  011031 (Mar 2014).

\bibitem{PhysRevLett.125.080501}
Xu, L., Liu, Z., Datta, A., Knee, G.~C., Lundeen, J.~S., Lu, Y.-q., and Zhang, L., ``Approaching quantum-limited metrology with imperfect detectors by using weak-value amplification,'' {\em Phys. Rev. Lett.}~{\bf 125},  080501 (Aug 2020).

\bibitem{PhysRevLett.126.220801}
Krafczyk, C., Jordan, A.~N., Goggin, M.~E., and Kwiat, P.~G., ``Enhanced weak-value amplification via photon recycling,'' {\em Phys. Rev. Lett.}~{\bf 126},  220801 (Jun 2021).

\bibitem{Liu:22}
Liu, Y., Zhang, Y., Xu, Z., Zhou, L., Zou, Y., Zhang, B., and Hu, Z., ``Ultra-low noise phase measurement of fiber optic sensors via weak value amplification,'' {\em Opt. Express}~{\bf 30},  18966--18977 (May 2022).

\bibitem{PhysRevLett.134.080802}
Huang, J.-H., Jordan, K.~M., Dada, A.~C., Hu, X.-Y., and Lundeen, J.~S., ``Enhancing interferometry using weak value amplification with real weak values,'' {\em Phys. Rev. Lett.}~{\bf 134},  080802 (Feb 2025).

\bibitem{PhysRevA.102.023701}
Li, Z., Xie, L., Ti, Q., Duan, P., Zhang, Z., and Ren, C., ``Increasing the dynamic range of weak measurement with two pointers,'' {\em Phys. Rev. A}~{\bf 102},  023701 (Aug 2020).

\bibitem{PhysRevA.100.012109}
Huang, J., Li, Y., Fang, C., Li, H., and Zeng, G., ``Toward ultrahigh sensitivity in weak-value amplification,'' {\em Phys. Rev. A}~{\bf 100},  012109 (Jul 2019).

\bibitem{PhysRevA.105.013718}
Huang, J.-H., He, F.-F., Duan, X.-Y., Wang, G.-J., and Hu, X.-Y., ``Modified weak-value-amplification technique for measuring a mirror's velocity based on the vernier effect,'' {\em Phys. Rev. A}~{\bf 105},  013718 (Jan 2022).

\bibitem{AWVA}
Huang, J.-H., Hu, X.-Y., Dada, A.~C., Lundeen, J.~S., Jordan, K.~M., Chen, H., and An, J., ``Autocorrelative weak-value amplification and simulating the protocol under strong gaussian noise,'' {\em Phys. Rev. A}~{\bf 106},  053704 (Nov 2022).

\bibitem{DVN/EOYJHO_2025}
Huang, J.-H., ``{Real-time analog circuit for auto-correlative weak-value amplification in the time domain: supplemental document},'' (2025).
\newblock https://doi.org/10.7910/DVN/EOYJHO, Harvard Dataverse,2025.

\bibitem{PhysRevLett.112.040406}
Ferrie, C. and Combes, J., ``Weak value amplification is suboptimal for estimation and detection,'' {\em Phys. Rev. Lett.}~{\bf 112},  040406 (Jan 2014).

\end{thebibliography}
\bibliographystyle{spiebib} 

\end{document}